\documentclass[a4paper,11pt]{article}
\usepackage{pos}
\newcommand{\mev}{\textrm{MeV}}
\newcommand{\gev}{\textrm{GeV}}
\newcommand{\fm}{\textrm{fm}}

\usepackage{subfig}

\title{Isovector Axial Form Factor of the Nucleon\\ from Lattice QCD}
\ShortTitle{Isovector Axial Form Factor of the Nucleon from Lattice QCD}

\author*[a]{Jonna Koponen}
\author[b,c]{Dalibor Djukanovic}
\author[a]{Georg von~Hippel}
\author[a,b]{Harvey B. Meyer}
\author[a]{Konstantin Ottnad}
\author[a]{Tobias Schulz}
\author[a,b]{Hartmut Wittig}

\affiliation[a]{PRISMA$^+$ Cluster of Excellence  \& Institut f\"ur Kernphysik,
Johannes Gutenberg-Universit\"at Mainz, D-55099 Mainz, Germany}
\affiliation[b]{Helmholtz-Institut Mainz, Johannes Gutenberg-Universit\"at Mainz,
D-55099 Mainz, Germany}
 \affiliation[c]{GSI Helmholtzzentrum für Schwerionenforschung, Darmstadt (Germany)}

\emailAdd{jkoponen@uni-mainz.de}

\abstract{The isovector axial form factor of the nucleon plays a key role in
interpreting data from long-baseline neutrino oscillation experiments.
We present a lattice QCD calculation of this form factor,
introducing a new method to directly extract its $z$-expansion from
lattice correlators.  Our final parameterization of the form factor,
which extends up to spacelike virtualities of $0.7\,{\rm GeV}^2$ 
with fully quantified uncertainties, agrees with previous lattice
calculations but is significantly less steep than neutrino-deuterium
scattering data suggests.}

\FullConference{%
  The 39th International Symposium on Lattice Field Theory (Lattice2022),\\
  8-13 August, 2022 \\
  Bonn, Germany 
}

\begin{document}
\maketitle

\section{Motivation}

The axial form factor of the nucleon $G_{\rm A}(Q^2)$ plays a central role in
understanding the quasi-elastic part of GeV-scale neutrino-nucleus cross sections.
These cross sections must be known with few-percent uncertainties~\cite{Ruso:2022qes}
to enable a reliable reconstruction of the incident neutrino energy
in the upcoming long-baseline neutrino oscillation experiments
DUNE~\cite{DUNE:2015lol} and T2HK~\cite{Hyper-Kamiokande:2018ofw}.
Lattice QCD determinations of $G_{\rm A}(Q^2)$~\cite{Meyer:2022mix} are crucial,
as currently available experimental measurements of the form factor fall short of the
required precision~\cite{Bernard:2001rs}.

Here we summarize the Mainz group's recent calculation of
$G_{\rm A}(Q^2)$ \cite{Djukanovic:2022wru} for momentum transfers up to
$0.7\,{\rm GeV}^2$ using lattice simulations with dynamical up, down and strange
quarks with an O($a$) improved Wilson fermion action. We employ a new analysis
method that simultaneously handles the issues of the excited-state contamination
and the description of the form factor's $Q^2$ dependence.

\section{Methodology}

The matrix elements of the local isovector axial current
$A^{a}_{\mu}(x) = \bar\psi \gamma_\mu\gamma_5\frac{\tau^a}{2}\psi$
between single-nucleon states are  parameterized by two form factors: the axial form
factor $G_{\rm A}(Q^2)$, and the induced pseudoscalar form factor $G_{\rm P}(Q^2)$. We
focus on calculating the axial form factor, which can be extracted from the current
component orthogonal to the momentum transfer.

The setup for the lattice calculation in this project \cite{Djukanovic:2022wru} is 
very similar to the one used in the case of the electromagnetic
form factors \cite{Djukanovic:2021cgp}. 
The nucleon two- and three-point functions are computed as
\begin{align}
\label{Eq:C2ptC3pt}
C_2(\vec{p},t) =& a^3\sum_{\vec{x}} e^{i\vec{p}\cdot\vec{x}} \Gamma_{\beta\alpha} \Big\langle \Psi^\alpha(t,\vec{x})\overline{\Psi}^\beta(0)\Big\rangle, \\
 C_{3}(\vec{q},t,t_s) =& -i a^6\sum_{\vec{x},\vec{y}}
e^{i\vec{q}\cdot\vec{y}} \Gamma_{\beta\alpha} \frac{\vec q\times \vec s}{|\vec q\times \vec s|^2}
\Big\langle \Psi^\alpha(t_s,\vec{x}) \vec q\times \vec A^{a=3}(t,\vec{y}) \overline{\Psi}^\beta(0)\Big\rangle,
\nonumber
\end{align}
where $t_s$ is the source-sink separation in the time direction, $\Psi^\alpha(\vec{x},t)$ denotes the proton interpolating
operator and $\Gamma = \frac{1}{2}(1 + \gamma_0)(1 + i \gamma_5 \vec s\cdot \vec\gamma)$ is the projection matrix.
We set $\vec s = \vec e_3$, aligning the nucleon spin along the $x_3$-axis.

\subsection{Summation method + $z$-expansion}

The accessible momentum transfers are discrete, and for
a given value of $\mathfrak{q}= {2\pi |\vec n|}/{L}$, we perform averages of the two-point functions 
over all spatial momenta $\vec q$ of the same norm $\mathfrak{q}$.
We then use the ratio
\begin{equation}
\label{Eq:ratio}
 R(\vec{q},t,t_s) \equiv \frac{C_{3}(\vec{q},t,t_s)}{\overline C_2(0,t_s)} \sqrt{ \frac{\overline C_2(|\vec{q}|,t_s-t)\overline C_2(0,t)\overline C_2(0,t_s)}{\overline C_2(0,t_s-t)\overline C_2(|\vec{q}|,t)\overline C_2(|\vec{q}|,t_s)}}
\end{equation}
to construct a momentum-averaged estimator for $ G_{\rm A}(Q^2)$,
\begin{equation}
G^{\rm eff}_{\rm A}(\mathfrak{q}; t,t_s) = \sqrt{\frac{2 E_{\mathfrak{q}}}{m+E_{\mathfrak{q}}}}
\sum_{|\vec q|=\mathfrak{q}} R(\vec{q},t,t_s) \Big/
\Big(\sum_{|\vec q|=\mathfrak{q}} 1 \Big).
\end{equation}
Here $m$ is the nucleon mass and $E_{\mathfrak{q}}$ its energy.
These effective form factors are then used to construct the summed insertion
\begin{equation}
\label{eq:summation}
S(\mathfrak{q},t_s) \equiv a \sum_{t=a}^{t_s-a}  G^{\rm eff}_{\rm A}(\mathfrak{q}; t,t_s)
\stackrel{t_s\to\infty}{=} b_0(\mathfrak{q}) + t_s   G_{\rm A}(Q^2) + \dots
\end{equation}
with momentum transfer $Q^2 = \mathfrak{q}^2 - (m-E_{\mathfrak{q}})^2$. The excited states, indicated by the ellipsis,
are of order $e^{-\Delta t_s}$ and $t_s e^{-\Delta t_s}$, with $\Delta$ the energy gap above
the single-nucleon state.

To extract the form factor $G_{\rm A}(Q^2)$, we parametrize it from the
outset via the $z$-expansion
\begin{equation}\label{Eq:zexp}
  G_{\rm A}(Q^2) = \sum^{n_{\rm max}}_{n=0} a_n z^n(Q^2),\quad
  z(Q^2)  = \frac{\sqrt{t_{\rm cut} + Q^2} - \sqrt{t_{\rm cut}} }{\sqrt{t_{\rm cut} + Q^2} + \sqrt{t_{\rm cut}}},
\end{equation}
fitting simultaneously the $\mathfrak{q}$ and $t_s$ dependence of $S(\mathfrak{q},t_s)$.
Here $t_{\rm cut} =  (3M_\pi^{\rm phys})^2$, and both the offsets $b_0(\mathfrak{q})$ (independent for each
$\mathfrak{q}$) and the coefficients $a_n$ are fit parameters.
We set $n_{\rm max} =2$ in our main analysis, and test that setting $n_{\rm max} =3$
gives consistent results. In the future, we will also consider parametrizing the $\mathfrak{q}$
dependence of $b_0(\mathfrak{q})$.

\subsection{Choice of source-sink separation}

\begin{figure}[!t]    
\centering
\includegraphics[width=0.98\columnwidth]{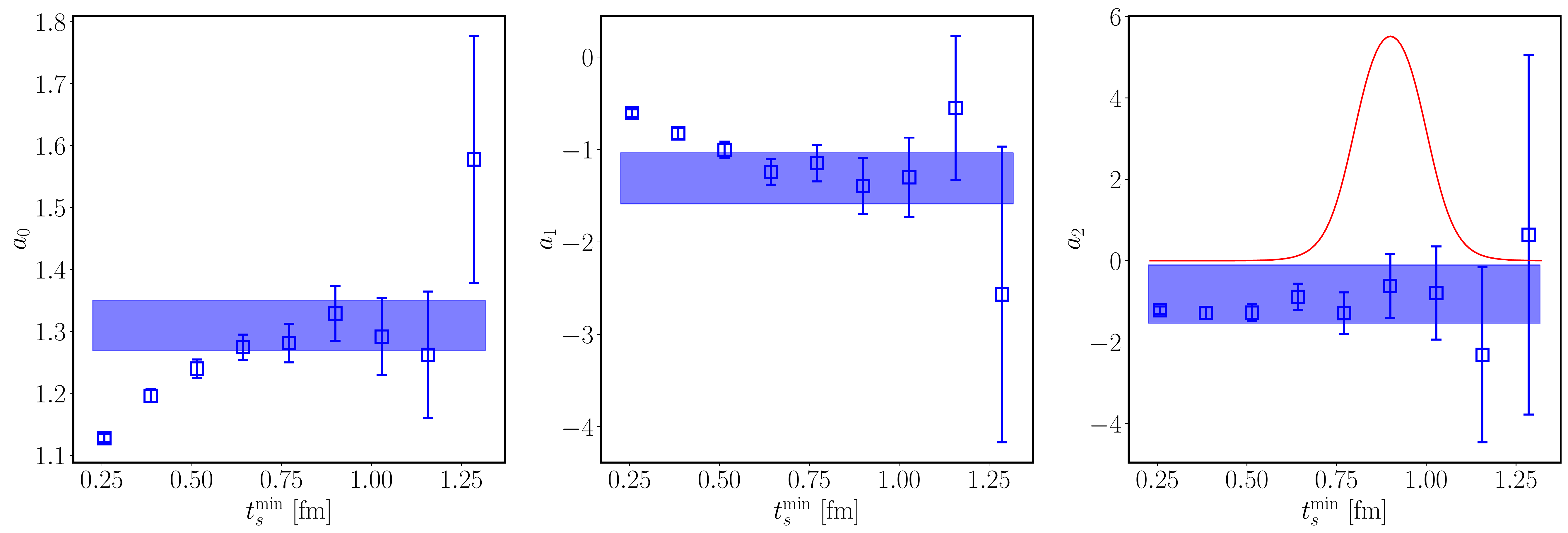}
\caption{Illustration of averaging over the minimum source-sink separation $t_{s}^{\mathrm{min}}$ in
  the summation method for the near-physical pion mass ensemble E250. We perform the $z$-expansion fits
  for each ensemble starting at different values of $t_{s}^{\mathrm{min}}$. The results for coefficients
  $a_0$, $a_1$ and $a_2$ are shown here as the blue squares. The bands represent the smooth-window
  averages over $t_{s}^{\mathrm{min}}$, and the solid red line shows the weight function 
  (arbitrarily normalized for visibility), which is applied to all three coefficients.}
\label{fig:tsep_min_average_E250}
\end{figure}

We perform fits to $S(\mathfrak{q},t_s)$ including all values of $t_s\geq t_s^{\textrm{min}}$, requiring that
at least two $t_s$ values enter the fit (and also requiring $n_\textrm{dof} > 0$).
At small values of $t_s$, contributions from excited states are expected to be significant, whereas at
large $t_s$ the signal-to-noise ratio becomes poor.
This leaves us with a relatively small window of starting values $t_s^{\textrm{min}}$ that can be used.
Rather than choosing a single $t_s^{\textrm{min}}$, we average the fit results $a_n(t_s^{\textrm{min}})$
over all values $t_s^{\textrm{min}}$, using a `smooth window' function $f(t_s^{\textrm{min}})$,
\begin{equation}
a_n = \sum_{t_s^{\textrm{min}}}  f(t_s^{\textrm{min}}) a_n(t_s^{\textrm{min}}) \Big/
\Big(\sum_{t_s^{\textrm{min}}} f(t_s^{\textrm{min}})\Big),\quad
f(\tau) = \tanh{\left(\frac{\tau -t_w^{\textrm{low}}}{\Delta t_w}\right)}-
\tanh{\left(\frac{\tau -t_w^{\textrm{up}}}{\Delta t_w}\right)},
\label{eq:smoothwindow}  
\end{equation}
as a weight factor.
We choose $t_w^{\textrm{low}}=0.8$~fm, $t_w^{\textrm{up}}=1.0$\;fm and $\Delta t_w=0.08$\;fm.
The average represents very well what one would identify as
a plateau in the fit results, as illustrated in Fig.~\ref{fig:tsep_min_average_E250}.

\section{Lattice ensembles}

We use a set of fourteen CLS $N_f=2+1$ ensembles~\cite{Bruno:2014jqa} that have been generated 
with non-perturbatively $\mathcal{O}(a)$-improved Wilson
fermions~\cite{Sheikholeslami:1985ij,Bulava:2013cta}
and the tree-level improved L\"uscher-Weisz gauge action~\cite{Luscher:1984xn}.
They cover the range of lattice spacings from $0.050$\;fm to $0.086$\;fm and pion masses
from about $350~\mev$ down to $130~\mev$.
Details of these ensembles, including the number of configurations, the number 
of measurements and the number of available source-sink separations $t_s$, are listed in Table~\ref{tab:ensembles}.
All ensembles used in this study have a fairly large volume, as indicated by $M_\pi L\gtrsim 4$. 

\begin{table}
 \centering
  \begin{tabular}{lcccccccccr}
   \hline\hline
   ID  & $\beta$ & $T/a$ & $L/a$ & $\!\!M_\pi/\mev$ & $M_\pi L$ & $\!\!M_N/\gev$ & $N_\mathrm{conf}$ & $N_\mathrm{meas}$ & $t_s~[\fm]$ & $N_{t_s}$ \\
   \hline\hline
   H102 & 3.40 &  96 & 32 & 354 & 4.96 & 1.103 & 2005 &  32080 & 0.35..1.47 & 14 \\ 
   H105 & 3.40 &  96 & 32 & 280 & 3.93 & 1.045 & 1027 &  49296 & 0.35..1.47 & 14 \\ 
   C101 & 3.40 &  96 & 48 & 225 & 4.73 & 0.980 & 2000 &  64000 & 0.35..1.47 & 14 \\ 
   N101 & 3.40 & 128 & 48 & 281 & 5.91 & 1.030 & 1596 &  51072 & 0.35..1.47 & 14 \\ 
   \hline                                                                    
   S400 & 3.46 & 128 & 32 & 350 & 4.33 & 1.130 & 2873 &  45968 & 0.31..1.53 &  9 \\ 
   N451 & 3.46 & 128 & 48 & 286 & 5.31 & 1.045 & 1011 & 129408 & 0.31..1.53 &  9 \\ 
   D450 & 3.46 & 128 & 64 & 216 & 5.35 & 0.978 &  500 &  64000 & 0.31..1.53 & 17 \\ 
   \hline                                                                    
   N203 & 3.55 & 128 & 48 & 346 & 5.41 & 1.112 & 1543 &  24688 & 0.26..1.41 & 10 \\ 
   N200 & 3.55 & 128 & 48 & 281 & 4.39 & 1.063 & 1712 &  20544 & 0.26..1.41 & 10 \\ 
   D200 & 3.55 & 128 & 64 & 203 & 4.22 & 0.966 & 2000 &  64000 & 0.26..1.41 & 10 \\ 
   E250 & 3.55 & 192 & 96 & 129 & 4.04 & 0.928 &  400 & 102400 & 0.26..1.41 & 10 \\ 
   \hline                                                                    
   N302 & 3.70 & 128 & 48 & 348 & 4.22 & 1.146 & 2201 &  35216 & 0.20..1.40 & 13 \\ 
   J303 & 3.70 & 192 & 64 & 260 & 4.19 & 1.048 & 1073 &  17168 & 0.20..1.40 & 13 \\ 
   E300 & 3.70 & 192 & 96 & 174 & 4.21 & 0.962 &  570 &  18240 & 0.20..1.40 & 13 \\ 
   \hline\hline
   \vspace*{0.1cm}
  \end{tabular}
  \caption{Overview of ensembles used in the study. The values $\beta=3.40$, $3.46$, $3.55$
    and $3.70$ correspond to lattice spacings $a\approx 0.086$, $0.076$, $0.064$ and $0.050~\fm$,
    respectively~\cite{Bruno:2016plf}. Columns $T/a$ and $L/a$ give the temporal and spatial
    size of the lattice, and $M_\pi$ and $M_N$ are the pion and nucleon masses. $N_\mathrm{conf}$
    is the number of configurations used for each ensemble, and in column $N_\mathrm{meas}$ we
    list the number of measurements done at the largest source-sink separation. $N_{t_s}$ is
    the number of available source-sink separations in the range listed in column $t_s$. For
  more details see~\cite{Djukanovic:2022wru}.}
 \label{tab:ensembles}
\end{table}

\section{Global fit}

To obtain the form factor at the physical point, the $a_n$ are extrapolated to the continuum
and interpolated to the physical pion mass, at which point the form factor may be evaluated
at any $q^2$ in the chosen expansion interval $[0,\;0.7\,{\rm GeV}^2]$.

For the extrapolation to the continuum, we include a term linear in $a^2$ for each of the coefficients.
For the extrapolation in pion mass, we use the following three ans\"atze:
\begin{enumerate}
\item
  Linear in $M_\pi^2$ for all coefficients $a_n$.
  
\item
  Again linear in $M_\pi^2$ for coefficients $a_1$ and $a_2$,
  and an extended ansatz containing a chiral logarithm for the zeroth coefficient:
  \begin{equation*}
    a_0 = g_a^{(0)}+g_a^{(1)}M_{\pi}^2+g_a^{(3)}M_{\pi}^3-g_a^{(2)}M_{\pi}^2\ln{\frac{M_{\pi}}{M_N}},
  \end{equation*}
  with $g_a^{(1)} = 4d_{16}-(g_a^{(0)})^3/(16\pi^2 F_\pi^2$, $g_a^{(2)} = g_a^{(0)}\left(1+2(g_a^{(0)})^2\right)/(8\pi^2 F_\pi^2)$,
  and $g_a^{(3)} = g_a^{(0)}\left(1+(g_a^{(0)})^2\right)/(8\pi F_\pi^2 M_N)-g_a^{(0)}\Delta_{c_3,c_4}/(6\pi F_\pi^2)$.
  Here  $M_N=938.92~\mev$ is the nucleon mass, $F_\pi=92.42~\mev$ is the pion decay constant~\cite{Schindler:2006it},
  and $\Delta_{c_3,c_4}=c_3-2c_4$ is a combination of low-energy constants $c_3$ and $c_4$. The free fit parameters for
  the zeroth coefficient's chiral extrapolation are $g_a^{(0)}$, $d_{16}$ and $\Delta_{c_3,c_4}$.

\item
  Same as ansatz~2, but including $M_\pi^3$ terms for coefficients $a_1$ and $a_2$.
  
\end{enumerate}
Note that, while the coefficients $a_n$ do not have common fit parameters, they are correlated
within an ensemble: these correlations are taken into account in the fits.

To check for possible finite-size effects (FSE), we include a term~\cite{Beane:2004rf}
  $\frac{M_{\pi}^2}{\sqrt{M_{\pi}L}}\mathrm{e}^{-M_{\pi}L}$
for the zeroth coefficient $a_0$  in some of the extrapolation fits. We do not observe a strong dependence
on the volume. 
Finite-size effects can also be inspected directly by comparing our results of the $z$-expansion fits on two ensembles,
H105 and N101, which differ only by their physical volume.
We find that
the coefficients $a_n$ agree well, confirming that finite-size effects are small at the current level of
precision.

We perform multiple extrapolations  using these different fit ans\"atze  with
pion mass cuts $M_{\pi} < M_{\pi}^{\rm cut}$ with $M_{\pi}^{\rm cut} =$ 300, 285, 265 and 250 MeV,
as well as dropping data from the coarsest lattice spacing, to get a handle on systematic effects.

\begin{figure}
\centering
\includegraphics[width=0.49\columnwidth]{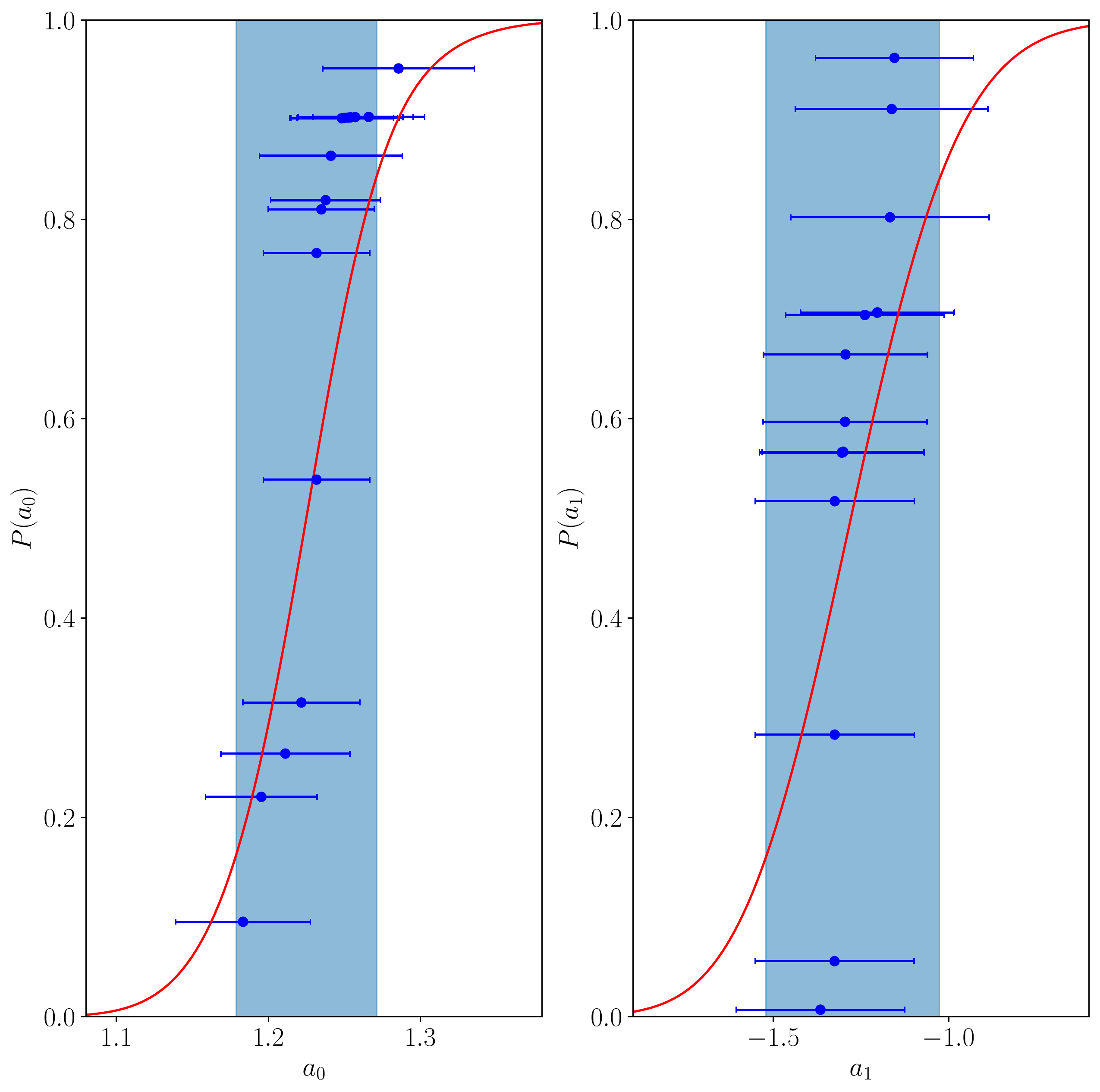}
\includegraphics[width=0.49\columnwidth]{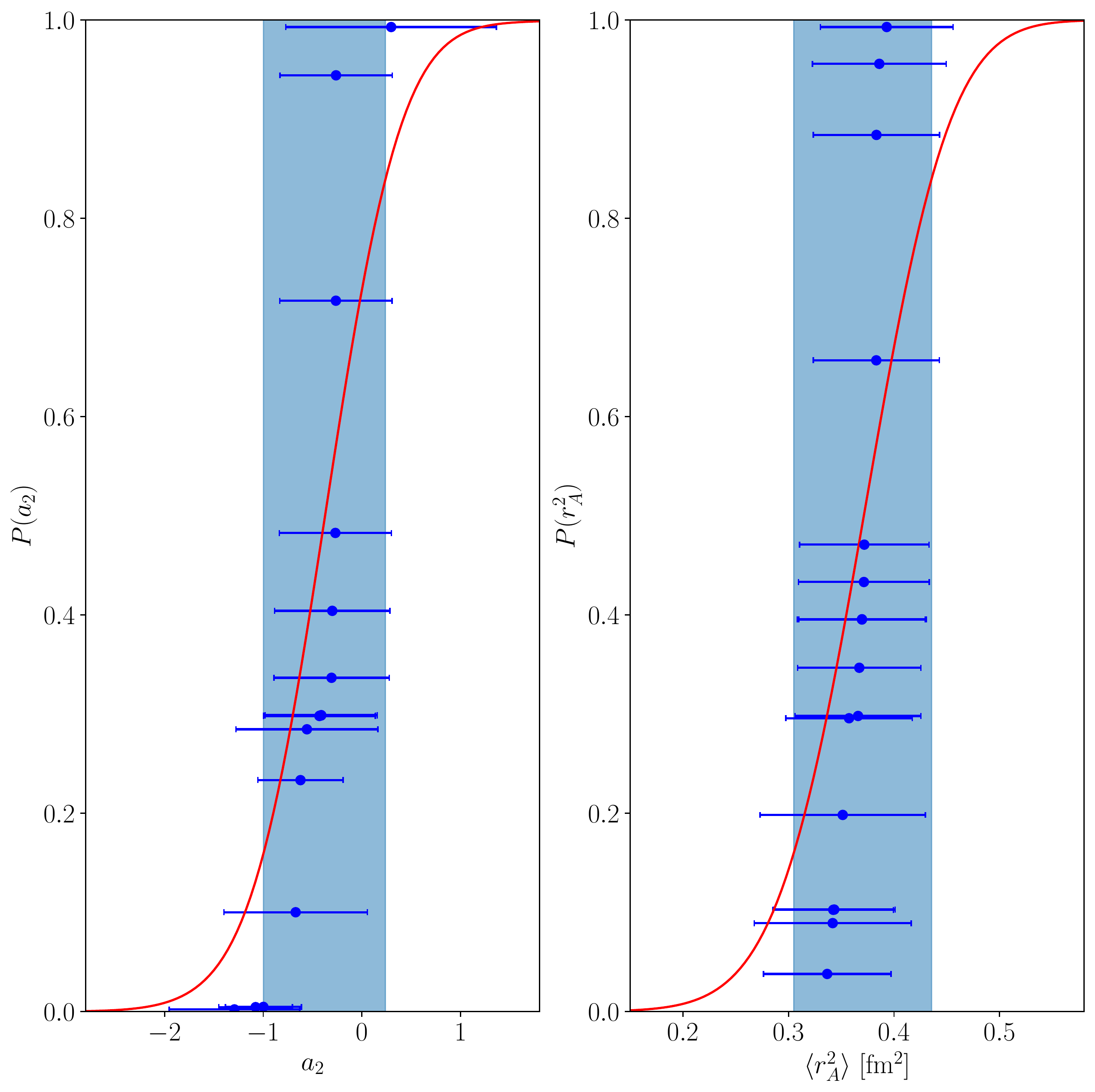}
\caption{AIC average and the corresponding cumulative distribution function for all coefficients
$a_0$, $a_1$, $a_2$, and the mean square radius $\langle r_A^2\rangle$ (in fm$^2$ units).}
\label{fig:a0_rA_AIC}
\end{figure}

\section{Model average (AIC)}

Since the different fit ans\"atze and cuts can be equally well motivated,
we perform a weighted average~\cite{Jay:2020jkz} over the resulting $a_n$. The Akaike Information
Criterion (AIC)~\cite{Akaike} is used to provide the weight to different analyses and to estimate
the systematic error associated with the variations of the global fit. Different variations of
the AIC weights have been used over the years, and we choose~\cite{Borsanyi:2020mff}
  $w^{\textrm{AIC}}=N\mathrm{e}^{-\frac{1}{2}\left(\chi^2+2n_{\textrm{par}}-n_{\textrm{data}}\right)}$,
where each fit is characterized by the minimum  $\chi^2$, the number of fit parameters $n_{\textrm{par}}$
and the number of data points $n_{\textrm{data}}$.
Here $N$ normalizes the weights so that their sum is 1.

The corresponding cumulative distribution functions $P(a_n)$ of the coefficients $a_n$ and of
the mean square radius $\langle r_A^2\rangle$ are shown in Fig.~\ref{fig:a0_rA_AIC}.
The full uncertainties, which are shown by the blue error bands, are determined by the
limits $P(a_n)=0.16$ and $P(a_n)=0.84$. The AIC procedure is also used to distinguish
the statistical and systematic components of the total uncertainty, following the
prescription proposed in~\cite{Borsanyi:2020mff}.

\section{Results}

Our results  for the coefficients of the $z$-expansion (Eq.~\eqref{Eq:zexp} with $t_{\rm cut}=(3M_{\pi^0})^2$)
of the nucleon axial form factor in the continuum and at the physical pion mass are
\begin{align}
  a_0 =& \quad\: 1.225 \pm 0.039 \textrm{ (stat)} \pm 0.025 \textrm{ (syst)}, \nonumber \\
  a_1 =& -1.274 \pm 0.237 \textrm{ (stat)} \pm 0.070 \textrm{ (syst)}, \label{eq:results_ai} \\
  a_2 =& -0.379 \pm 0.592 \textrm{ (stat)} \pm 0.179 \textrm{ (syst)}, \nonumber   
\end{align}  
with a correlation matrix
\begin{equation}
M_{\textrm{corr}} = \left(\!
\begin{array}{rrr}
 1.00000 & -0.67758 &  0.61681 \\
-0.67758 &  1.00000 & -0.91219 \\
 0.61681 & -0.91219 &  1.00000
\end{array} \,\right).
\end{equation}

\begin{figure} 
\subfloat[][]{
\includegraphics[trim={0mm -2.5mm 0mm 2.3mm},clip,width=0.389\columnwidth]{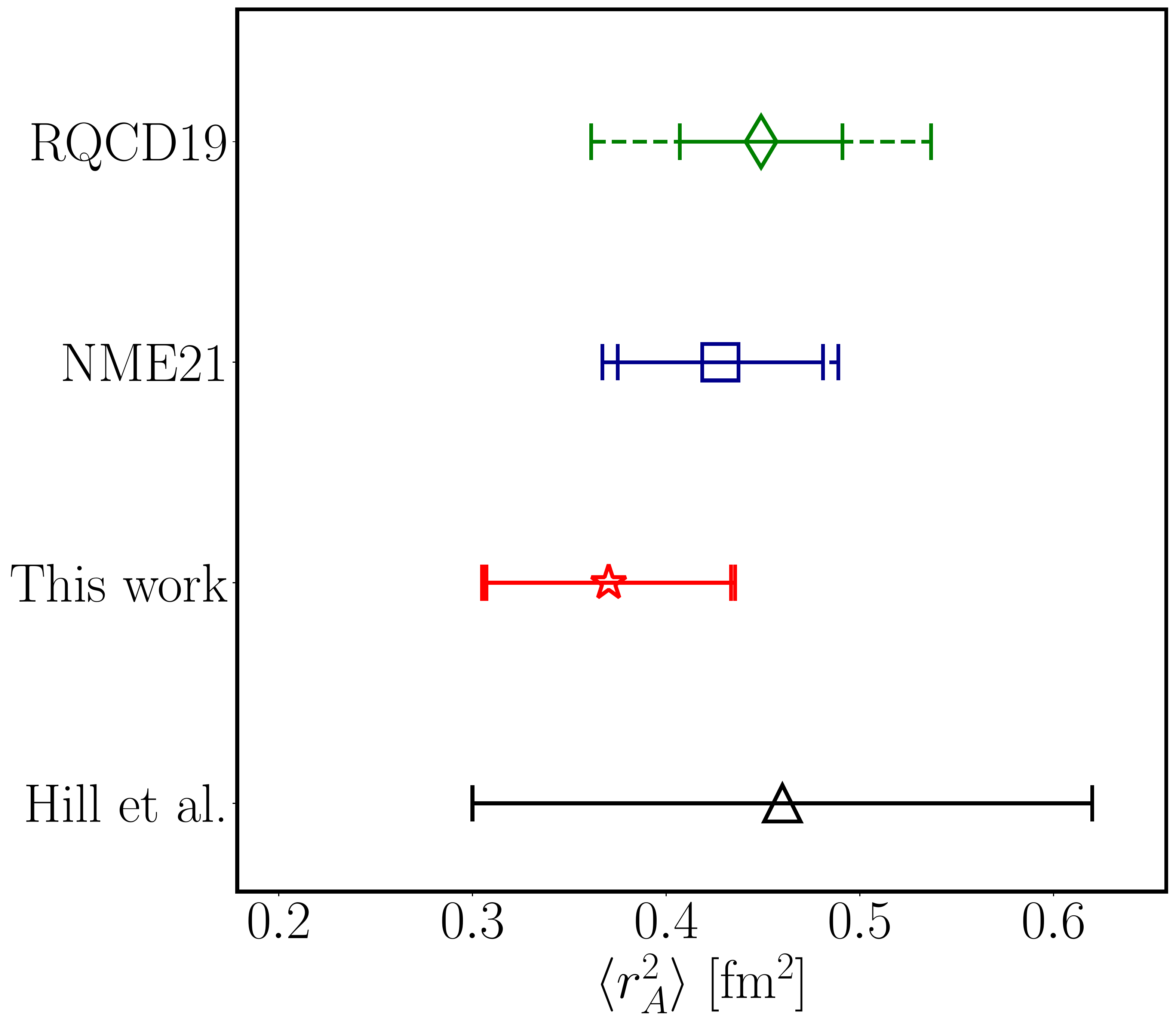}}
\subfloat[][]{
\includegraphics[trim={0mm 2mm 2mm 2.4mm},clip,width=0.598\columnwidth]{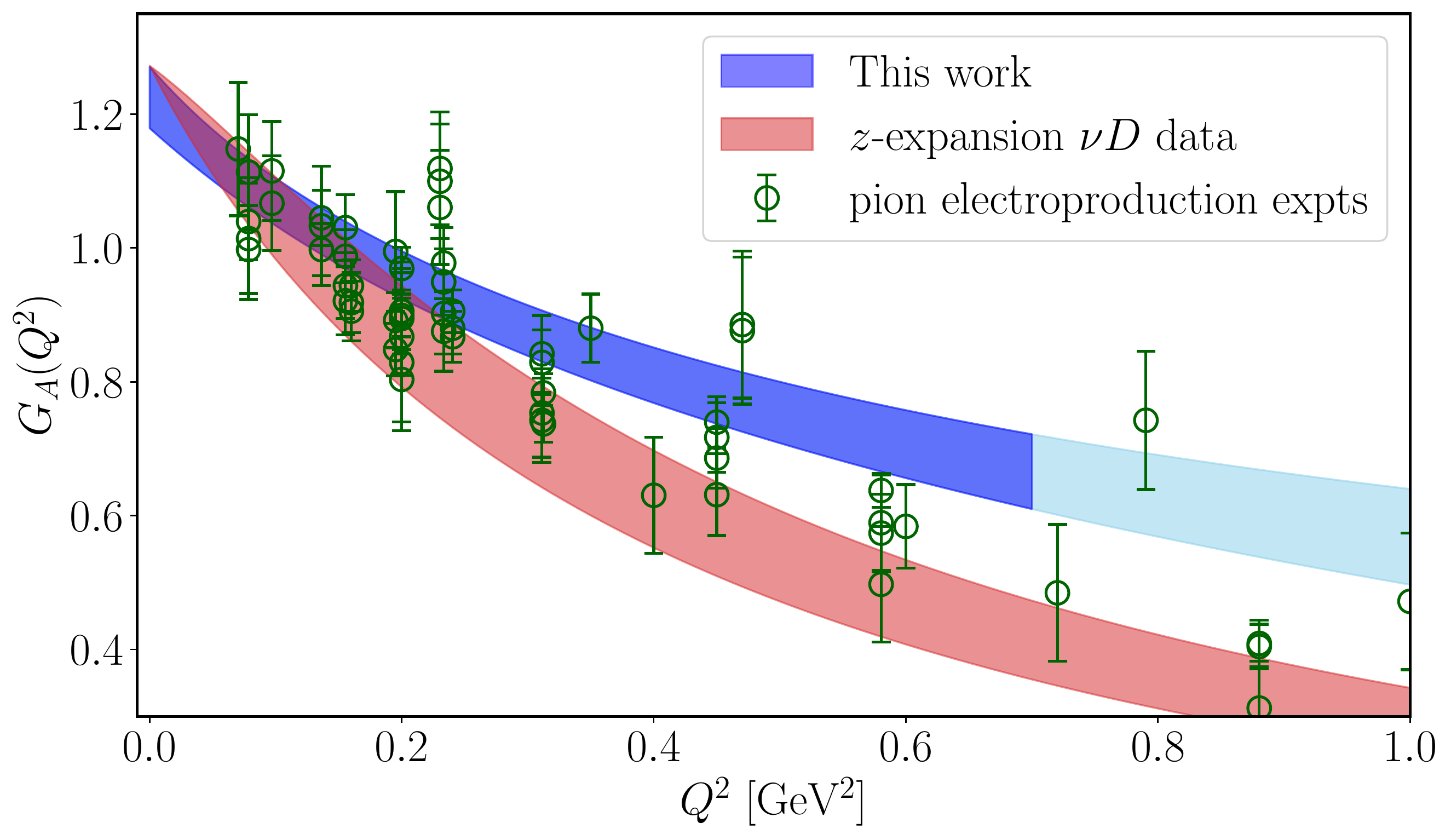}}
\caption{(a): Comparison of lattice determinations of the mean square radius
$\langle r_\text{A}^2\rangle$, from Refs.~\cite{Park:2021ypf}
(NME21) and \cite{RQCD:2019jai} (RQCD20), which have full error budget (chiral and
  continuum extrapolation). 
  The point labeled Hill et al. is  an average of the values obtained from $z$-expansion
fits to neutrino scattering and muon capture \cite{Hill:2017wgb}.
The smaller error bars with solid lines show the statistical errors, whereas
the wider error bars with dashed lines show the total errors (including systematic
uncertainties).
(b): Comparing our result for the axial form factor to data from pion electroproduction
experiments~\cite{Bernard:2001rs} (normalized by the PDG value for the axial charge~\cite{PDG2022})
and to a $z$-expansion fit to neutrino-Deuterium scattering
  data~\cite{Meyer:2016oeg}. There is a clear tension between the lattice QCD result and the
  $z$-expansion extracted from deuterium bubble chamber data, especially at large $Q^2$. The
  darker blue error band highlights the $Q^2$ range of our lattice data.
}
\label{fig:rAsq_and_FF}
\end{figure}

This leads to the mean square radius
$\langle r^2_A \rangle = (0.370 \pm 0.063 \textrm{ (stat)} \pm 0.016 \textrm{ (syst)} )~\fm^2$,
which is in good agreement with other lattice QCD determinations
-- see Fig.~\ref{fig:rAsq_and_FF}. The comparison features only lattice calculations with a
full error budget, including a continuum extrapolation: The NME21 result is from~\cite{Park:2021ypf},
and the RQCD20 result is from~\cite{RQCD:2019jai}.
Both studies parameterize the $Q^2$ dependence of the form factor using a $z$-expansion (RQCD
also use a dipole ansatz as an alternative parameterization, the result of which is not shown
in the figure). Other lattice
calculations~\cite{Jang:2019vkm,Alexandrou:2020okk,Shintani:2018ozy,Ishikawa:2021eut,Hasan:2017wwt}
also exist with partial error budgets.
For comparison, we show the average of the values obtained from $z$-expansion
fits to neutrino scattering and muon capture measurements~\cite{Hill:2017wgb}. Our result also agrees well
with the earlier two-flavour calculation by the Mainz group~\cite{Capitani:2017qpc}, and with
a more recent analysis~\cite{Schulz:2021kwz} by the same group that has
been obtained via the conventional two-step process of first determining the form factor at discrete
$Q^2$ values and then parameterizing it using a $z$-expansion.

The axial charge $g_A= a_0$ is in good agreement with our previous determination~\cite{Harris:2019bih}
based on forward nucleon matrix elements only. Since that method tends to yield more precise results
for a given data set,
we do not view the present determination of $g_A$ as superseding that of Ref.~\cite{Harris:2019bih},
and merely perform the comparison as a consistency check.

Finally, we compare our result for the axial form factor to data
from pion electroproduction experiments~\cite{Bernard:2001rs} and to a $z$-expansion fit
to neutrino-Deuterium scattering data~\cite{Meyer:2016oeg} in Fig.~\ref{fig:rAsq_and_FF}.
Our result agrees well with other lattice QCD calculations, as can be seen by comparing
this figure to Fig.~3 in the recent review~\cite{Meyer:2022mix}. However, there is a tension with the
axial form factor extracted from experimental deuterium bubble chamber data~\cite{Meyer:2016oeg},
especially at large $Q^2$. According to the authors of the Snowmass White Paper on
Neutrino Scattering Measurements~\cite{Alvarez-Ruso:2022ctb}, this discrepancy suggests
that a 30-40\% increase would be needed in the nucleon quasielastic cross section
for the two results to match. They also note that recent high-statistics data on nuclear targets
cannot directly resolve such discrepancies due to nuclear modeling uncertainties.

\section{Summary and conclusions}

In this report, we have given a summary of the Mainz group's recent publication~\cite{Djukanovic:2022wru},
which introduced a new method to extract the axial form factor of the nucleon.
It combines two well-known methods into one analysis step:
the summation method, which ensures that excited-state effects
are sufficiently suppressed, and the $z$-expansion, which provides the parameterization of
the $Q^2$ dependence of the form factor. Our main results are the coefficients of the
$z$-expansion, given in Eq.~(\ref{eq:results_ai}).
Systematic effects are included through AIC weighted average, which also
provides the break-up into statistical and systematic uncertainties and
the correlations among the coefficients.

We observe good agreement with other lattice QCD determinations of the
axial form factor, which confirms and strengthens the tension with the shape of the
form factor extracted from deuterium bubble chamber data. Comparing our
result for $a_0\equiv G_{\rm A}(0)$  to the Particle Data
Group (PDG) value for the axial charge, $g_A = 1.2754(13)$ \cite{PDG2022},
which can be viewed as a benchmark, we find agreement at the
$1.1\,\sigma$ level. Also, previously, a good overall agreement was found for
the isovector vector form factors~\cite{Djukanovic:2021cgp} with
phenomenological determinations, which are far more precise than in
the axial-vector case. This all adds confidence to the finding that
the nucleon axial form factor falls off less steeply 
than previously thought.

\acknowledgments{We thank Tim Harris, who was involved in the early
  stages of this project. %~\cite{Brandt:2017vgl}.
  This work was supported in part by the European Research Council (ERC) under the
  European Union’s Horizon 2020 research and innovation program
  through Grant Agreement No.\ 771971-SIMDAMA and by the Deutsche
  Forschungsgemeinschaft (DFG) through the Collaborative Research
  Center SFB~1044 ``The low-energy frontier of the Standard Model'',
  under grant HI~2048/1-2 (Project No.\ 399400745) and in the Cluster
  of Excellence “Precision Physics, Fundamental Interactions and
  Structure of Matter” (PRISMA+ EXC 2118/1) funded by the DFG within
  the German Excellence strategy (Project ID 39083149).
  
  Calculations for this project were partly performed on the HPC
  clusters ``Clover'' and ``HIMster2'' at the Helmholtz Institute Mainz,
  and ``Mogon 2'' at Johannes Gutenberg-Universit\"at Mainz.
  The authors gratefully acknowledge the Gauss Centre for Supercomputing e.V.
  (www.gauss-centre.eu) for funding this project by providing computing
  time on the GCS Supercomputer systems JUQUEEN and JUWELS at J\"ulich
  Supercomputing Centre (JSC) via grants NUCSTRUCLFL, CHMZ23, CHMZ21 and CHMZ36 (the latter
  through the John von Neumann Institute for Computing (NIC)), as well as
  on the GCS Supercomputer HAZELHEN at H\"ochstleistungsrechenzentrum
  Stuttgart (www.hlrs.de) under project GCS-HQCD.
  
Our programs use the QDP++ library~\cite{Edwards:2004sx} and deflated SAP+GCR
solver from the openQCD package~\cite{Luscher:2012av}, while the contractions
have been explicitly checked using~\cite{Djukanovic:2016spv}. We are grateful to
our colleagues in the CLS initiative for sharing the gauge field configurations.
}

\bibliographystyle{JHEP}
\bibliography{references.bib} 

\end{document}